\documentclass{article}
\usepackage{amsmath}
\usepackage{amsfonts}
\usepackage{amssymb}
\usepackage[usenames]{color}
\usepackage{amsthm}

\def\l{\lambda}
\def\p{\partial}
\def\e{\mathrm{e}}
\newtheorem{prop}{Proposition}

\theoremstyle{remark}
\newtheorem*{remark}{Remark}

\newcommand{\wt}{\widetilde}
\newcommand{\be}{\begin{equation}}
\newcommand{\ee}{\end{equation}}
\newcommand{\bea}{\begin{eqnarray}}
\newcommand{\eea}{\end{eqnarray}}
\newcommand{\beaa}{\begin{eqnarray*}}
\newcommand{\eeaa}{\end{eqnarray*}}

\newcommand{\nn}{\nonumber}
\renewcommand{\d}{\mathrm{d}}
\newcommand{\Red}{\textcolor{red}}
\usepackage{authblk}
\begin{document}
\title
{Dispersionless integrable systems and the Bogomolny
equations on an Einstein-Weyl geometry background
\thanks
{This research was performed in the framework 
of State Assignment Topic 0033-2019-0006 
(Integrable systems
of Mathematical Physics).}
}
\author{L.V. Bogdanov \thanks{leonid@itp.ac.ru}}
\affil{Landau Institute for Theoretical Physics, RAS, 
Moscow, Russia}
\date{}
\maketitle
\begin{abstract}
We derive a dispersionless integrable system describing
a local form of a general three-dimensional
Einstein-Weyl geometry 
with an Euclidean (positive) signature, construct its 
matrix extension and demonstrate that it leads
to the Bogomolny equations for a non-abelian monopole
on an Einstein-Weyl geometry background.
The corresponding
dispersionless integrable hierarchy,
its matrix extension and the dressing scheme 
are also considered.
\end{abstract}

\textbf{Keywords:} 
Dispersionless integrable system,
Einstein-Weyl geometry,
\\
Bogomolny equations,
Yang-Mills-Higgs equations
\section{Introduction}
The present paper is a continuation of the work 
\cite{LVB20}, where an integrable matrix extension 
of the Manakov-Santini system \cite{MS06}, \cite{MS07}
was introduced and it was demonstrated that it describes
a (2+1)-dimensional integrable chiral model 
in an Einstein-Weyl space. 
Since the Manakov-Santini system corresponds to
a local form of a general Lorentzian 
Einstein-Weyl geometry \cite{DFK14}
(in the complex case, to a general
complex analytic one), its integrable matrix 
extension gives a local description 
of a general case of the integrable chiral
model model on an Einstein-Weyl geometry background.

The structure of the Bogomolny equations 
for a non-abelian monopole
(see, e.g., \cite{Dun}) is analogous
to the Yang-Mills-Higgs equations,
representing a reduction 
of the self-dual Yang-Mills equations
and leading to the integrable
chiral model \cite{Ward},
but, in contrast to it,
they are written in Euclidean,
not in Lorentzian three-dimensional space.
To consider these equations on an integrable
geometric background, one needs a local
description of an Einstein-Weyl geometry
with a positive (Euclidean) signature.

It is problematic to perform a reduction
to real case with Euclidean signature 
for the complex Manakov-Santini system
because of asymmetry with respect to
independent variables. A more suitable candidate
is a local description of an Einstein-Weyl
geometry using a generalisation of a
dispersionless 
(2+1)-dimensional Toda equation (d2DTL) \cite{LVB10}
\bea
&&
({e}^{-\phi})_{tt}=m_t\phi_{xy}-m_x\phi_{ty},
\nn\\
&&
m_{tt}{e}^{-\phi}=m_{ty}m_x-m_{xy}m_t,
\label{gen2DTL}
\eea
with the Lax pair
\bea
&&
\partial_x\mathbf{\Psi}=\left((\l+ \frac{m_x}{m_t})\partial_t - 
(\phi_t \frac{m_x}{m_t}-\phi_x)\l\partial_\l\right)\mathbf{\Psi},
\nn\\
&&
\partial_y\mathbf{\Psi}=\left(-\frac{1}{\l}\frac{{e}^{-\phi}}{m_t}\partial_t -
\frac{1}{\l}\frac{({e}^{-\phi})_t}{m_t}\l\partial_\l\right)\mathbf{\Psi}.
\label{Laxgen2DTL}
\eea
For $m=t$ system (\ref{gen2DTL}) reduces to
dispersionless 
(2+1)-dimensional Toda equation (d2DTL)
\bea
({e}^{-\phi})_{tt}=\phi_{xy},
\label{d2DTL}
\eea
A conformal structure of an Einstein-Weyl 
geometry is defined by the symbol
of linearization
of the equation representing a 
symmetric bivector (metric with upper indices)
\cite{FK14}, for equation (\ref{d2DTL}) it is
\beaa
h=\partial_x\partial_y + \e^{-\phi}\partial_t^2
\eeaa
In this case it is easy to obtain 
a real metric with an Euclidean signature
substituting
$x=z$, $y=\Bar z$,
where $z=x^1 + \mathrm{i} x^2$, to equation
(\ref{d2DTL}), which gives
\bea
({e}^{-\phi})_{tt}=\phi_{z\Bar z},
\label{d2DTLz}
\eea
and the corresponding conformal structure
\beaa
h=\partial_z\partial_{\Bar z} + \e^{-\phi}\partial_t^2
\eeaa
has an Euclidean signature in terms of real variables
$x^1$, $x^2$, $t$. 
Lax pair for this equation can be written
in a symmetric form
\bea
&&
\partial_{z}\mathbf\Psi=L_1\mathbf\Psi,\quad L_1
=
\lambda \e^{\varphi} \partial_t  
- (\varphi_z + \lambda \e^\varphi \varphi_\tau)
\l\partial_\lambda,
\nn\\
&&
\partial_{\bar z}\mathbf\Psi=L_2\mathbf\Psi,
\quad L_2
=
-\frac{1}{\lambda} \e^\varphi\partial_t 
+ 
(\varphi_{\Bar z}-\frac{1}{\l}\e^\varphi \varphi_\tau)
\l\partial_\lambda,
\label{LaxToda}
\eea
here $\varphi=-\frac{1}{2}\phi$. 
For real $\varphi$ we have a symmetry
$L_2(\lambda)=\bar L_1(-\bar\lambda^{-1})$.
Just this symmetry is responsible
for real Euclidean reduction of 
the conformal structure,
it is present for higher hierarchy
operators also.

For system (\ref{gen2DTL}) similar reduction 
does not work due to the asymmetry with respect to
variables $x$, $y$. However, it was demonstrated
in \cite{LVB10} that using a rather simple 
complex transformation involving a dependent 
variable $m$ and an independent variable $t$
it is possible to bring system (\ref{gen2DTL})
and its Lax pair to a symmetric form, providing
real Euclidean reduction. This transformation
looks natural in terms of the hierarchy
connected with system (\ref{gen2DTL}).

The goal of the present work is to construct
a symmetric analogue of
equations (\ref{gen2DTL}) and their
integrable matrix extension 
and a corresponding local form 
of an Einstein-Weyl geometry  
with an Euclidean signature.
We will demonstrate that the matrix
extension describes the Bogomolny equations
on an Einstein-Weyl geometry background.

We will take system (\ref{gen2DTL}) as a basic system,
corresponding Einstein-Weyl geometry was found
in the work \cite{FK14}. First we will demonstrate
that matrix extension of this system, constructed
in the work \cite{LVB20}, gives
the Yang-Mills-Higgs equations \cite{Dun} 
on the Lorentzian
Einstein-Weyl geometry background and, after fixing
the gauge, an integrable chiral model
(this conjecture was made but not proved in the work 
\cite{LVB20}). Then using a complex transformation
we will proceed to a symmetric analogue of
equations  (\ref{gen2DTL}) and the Lax pair,
construct their integrable matrix extension
and demonstrate that it represents the 
Bogomolny equations on an Einstein-Weyl geometry
background (the Yang=Mills-Higgs equations \cite{Dun}
on an Euclidean Einstein-Weyl geometry
background). In the final part of the work
we will build the hierarchy for symmetric analogue
of equations (\ref{gen2DTL}) and its matrix extension and outline
the dressing method scheme.

Twistor integrability of geometric structures
related to our paper was established
in the works
\cite{Penrose}, \cite{Atiah}, \cite{Hitchin}.
Integrable background geometries for different
reductions of the self-dual Yang-Mills equations
connected with dispersionless integrable systems
are considered in the paper \cite{Calderbank}.
The scheme of integrable matrix extension
of dispersionless integrable systems
was proposed in the work \cite{LVB17}.

\section{A matrix extension of system (\ref{gen2DTL})}
\subsection{An Einstein-Weyl geometry}
The Einstein-Weyl geometry corresponding to system
(\ref{gen2DTL}) was found in the work
\cite{FK14}.
A symbol of formal linearisation of system
(\ref{gen2DTL})
leads to  a symmetric bivector of a conformal structure
\be
h\sim 
e^{-\phi}\partial_t^2-
m_x\partial_y\partial_t+m_t\partial_x\partial_y,
\label{Weyl0}
\ee
which corresponds to the metric
\be
g=(m_xdx+m_tdt)^2+4e^{-\phi}m_tdxdy,
\label{Weyl1}
\ee
satisfying together with a differential form
\be
\omega=\left(\frac{m_{tt}}{m_t^{2}}-
2\frac{\phi_t}{m_t}\right)(m_x\,dx+m_t\,dt)
+ 2\frac{m_{yt}}{m_t}\,dy
\label{Weyl2}
\ee
Einstein-Weyl equations.

Let us remind (see 
\cite{FK14}, \cite{PT93} for more detail), 
that a Weyl space is a manifold with
a conformal structure
$[g]$ and a symmetric connection
${{D}}$ consistent with $[g]$ in a sense that
for every $g\in [g]$
\bea
{{D}}g=\omega\otimes g
\label{Weyl}
\eea
for some differential form $\omega$ (the connection
preserves a conformal class). The change of a metric 
representing a conformal structure
$g\rightarrow f g$,
where the function $f\neq 0$, corresponds 
to the transformation
$\omega\rightarrow \omega + \,\d \ln f$.
For the case of closed form 
$\omega$ 
it is possible locally to choose such a conformal
gauge that the connection
${{D}}$ coincides with a metric connection. 
In a general case the form
$\omega$ defines a difference of the connection
${{D}}$ 
from a metric Levi-Civita connection
$\nabla$,
explicit formulae for Christoffel symbols
of the connection 
${{D}}$ 
in terms of the form 
$\omega$ and Levi-Civita connection
for a metric $g$ are given, e.g.,
in the work 
\cite{PT93},
\beaa
D_i V^j=\nabla_i V^j + \gamma^{j}_{~ik}V^k,
\eeaa
here
\beaa
\gamma^{j}_{~ik}=-\tfrac{1}{2}
(\delta^j_i\omega_k+\delta^j_k\omega_i
-g_{ik}g^{jm}\omega_m).
\eeaa

Einstein-Weyl spaces are defined by the condition that
the trace-free part of the symmetrised Ricci tensor of
the connection  
${{D}}$ vanishes (Einstein equations),
which together with relation 
(\ref{Weyl}) 
constitute Einstein-Weyl equations system,
in a coordinate form
\bea
D_i g_{ij}=\omega_i g_{ij},
\quad R_{(ij)}=\Lambda g_{ij},
\label{EW0}
\eea
here $\Lambda$ is some function.
Einstein-Weyl equations are correctly
defined for arbitrary manifold dimension not
less than three,  but the most interesting case
is three-dimensional when they are integrable 
\cite{Hitchin}. 
We cosider only three-dimensional Einstein-Weyl
geometry in the present work.

\subsection{A matrix extension
and the Yang-Mills-Higgs equations
}
In the work \cite{LVB20} we introduced a matrix extension
of  Lax pair
(\ref{Laxgen2DTL})
\be
\begin{split}
L_1&=\partial_x-(\lambda+ \frac{m_x}{m_t})\partial_t + 
\lambda(\phi_t \frac{m_x}{m_t}-\phi_x)\partial_\lambda
+A
\\
L_2&=\partial_y+\frac{1}{\lambda}
\frac{{\e}^{-\phi}}{m_t}\partial_t 
+
\frac{(\mathrm{e}^{-\phi})_t}{m_t}\partial_\lambda
+\frac{1}{\lambda}B,
\end{split}
\label{Laxmatrix}
\ee
where $A$, $B$ are matrices (Lie algebra elements).
Commutativity conditions for the matrix extension lead
simultaneously to system 
(\ref{gen2DTL}) and matrix equations on the background 
of solutions of this system
\be
\begin{split}
&\p_t B +\p_y A=0,
\\
&(\partial_x-\frac{m_x}{m_t}\partial_t)B
- \frac{\mathrm{e}^{-\phi}}{m_t}\partial_t A
-(\phi_t \frac{m_x}{m_t}-\phi_x)B + [A,B]=0,
\end{split}
\label{TodaM} 
\ee
or, in terms of potential $K$,
$A=K_t$, $B=-K_y$,
\be
\e^{-\phi}K_{tt} - {m_x}K_{yt} + {m_t}K_{xy}
-(\phi_t {m_x}-\phi_x {m_t})K_y -{m_t}[K_y,K_t]=0.
\ee
This equation on the trivial background
$m=t$, $\phi=1$ coincides (up to a change of variables)
with an integrable chiral model
\cite{Dun}.

We would like to demonstrate 
that the matrix extension of system (\ref{gen2DTL})
corresponds to the Yang-Mills-Higgs equations on 
an Einstein-Weyl geometry (\ref{Weyl1}), (\ref{Weyl2})
background, similar to the result obtained in the 
work \cite{LVB20} for the matrix extension
of the Manakov-Santini system.
The Yang-Mills-Higgs equations on 
an Einstein-Weyl geometry background read
\bea
\mathrm{D}\Phi+\tfrac{1}{2}\omega \Phi=*{F},
\label{Ward}
\eea
where  a 2-form
${F}=\d{A}+{A}\wedge{A}$ (
curvature of the connection,
gauge field intensity),
${A}$ is a gauge field (potential)
representing a 1-form with the values 
in some (matrix) Lie algebra, $\Phi$ 
is a function taking values in the Lie
algebra (Higgs field, \cite{Dun}), 
$\mathrm{D}\Phi=\d\Phi+[{A},\Phi]$. 
The form $\omega$ together with the metric
$g$ satisfy Einstein-Weyl equations
(\ref{EW0}).
For the Minkowski metric 
equation (\ref{Ward})
coincides with the Yang-Mills-Higgs
system introduced by Ward \cite{Ward},
leading to an integrable chiral model,
in the Euclidean case it gives 
the Bogomolny equations (see, e.g., \cite{Dun}). 
Equation (\ref{Ward}) is invariant under the change
of a conformal gauge of the Einstein-Weyl geometry 
$g\rightarrow f g$,
$\omega\rightarrow \omega + \,\d \ln f$,
$\Phi\rightarrow f^{-\frac{1}{2}}\Phi$.

We will introduce a special basis of vector fields
and a dual basis of differential forms in which
metric (\ref{Weyl1}) takes a simple form, and will
consider components of equation (\ref{Ward})
in this basis with the use of a standard formula
\bea
{F}({u},{v})=\nabla_{u}\nabla_{v}
-\nabla_{v}\nabla_{u}-
\nabla_{[{u},{v}]}
\label{F}
\eea
which is valid for arbitrary vector fields
${u}$, ${v}$.

Let us introduce a basis of vector fields
\bea
\mathbf{e}_1=\partial_x-\frac{m_x}{m_t}\partial_{t},
\quad
\mathbf{e}_2=\partial_{y},
\quad
\mathbf{e}_3=\partial_t,
\label{basisv}
\eea
in which symmetric bivector of conformal structure
(\ref{Weyl0}) is represented as
\bea
h=
\e^{\phi}m_t
\mathbf{e}_1\cdot \mathbf{e}_2 
+ (\mathbf{e}_3)^2.
\label{h0}
\eea
The dual basis of 1-forms reads
\bea
\mathbf{e}^1=\d x,
\quad
\mathbf{e}^2=\d y,
\quad
\mathbf{e}^3=
(\d {t} + \frac{m_x}{m_t}\d {x}),
\label{basisf}
\eea
and for metric
(\ref{Weyl1}) we obtain
\bea
g=
4\frac{\e^{-\phi}}{m_t}
\mathbf{e}^1\cdot \mathbf{e}^2 
+ (\mathbf{e}_3)^2
\label{metric0}
\eea
(up to a conformal factor $m_t^{-2}$).

Let us consider gauge covariant vector fields
$\nabla_i=\nabla_{\mathbf{e}_i}=\mathbf{e}_i + A_i$.
In terms of these fields Lax pair
(\ref{Laxmatrix}) 
takes the form
\bea
L_1=\nabla_1 - 
\lambda (\nabla_{3} + \Phi)
+
(\phi_t \frac{m_x}{m_t}-\phi_x)
\l\partial_\lambda,
\nn\\
L_2=\nabla_2 + 
\frac{1}{\lambda}\frac{{\e}^{-\phi}}{m_t}
(\nabla_{3} - \Phi)
+
\frac{1}{\lambda}
\frac{(\mathrm{e}^{-\phi})_t}{m_t}
\l\partial_\lambda
\label{Laxgauge}
\eea
Operators (\ref{Laxmatrix}) 
correspond to a special gauge compatible
with commutativity conditions.
We will consider matrix extension
(\ref{Laxgauge})
without fixing a gauge.
From the commutativity equations we obtain
\bea
&&
-\frac{1}{2}{\e}^{\phi}{m_t}
[\nabla_1,\nabla_2]
=
[\nabla_3,\Phi]
+
\frac{1}{2}
\frac{m_{tt}}{m_t}
\nabla_3
-
\frac{1}{2}
\left(
2\phi_t+\frac{m_{tt}}{m_t}
\right)
\Phi,
\label{comm1}
\\
&&
[\nabla_2,\nabla_3+\Phi]=0,
\label{comm2}
\\
&&
[\nabla_1,\nabla_3-\Phi]
=
\left(\frac{m_x}{m_t}\right)_t
(\nabla_3-\Phi)
\label{comm3}
\eea
We have omitted commutativity equations corresponding
to the terms containing a derivative over spectral 
variable $\p_\lambda$. These equations do not contain 
a matrix part and they reduce to the first equation of
system (\ref{gen2DTL}).
The second equation of this system 
arises from the scalar (corresponding to commutation
of vector fields) part of relation
(\ref{comm1}). 
Relation (\ref{comm2}) 
is responsible for the existence of a gauge
in which system 
(\ref{Laxgauge}) 
reduces to a form (\ref{Laxmatrix}).

According to formula (\ref{F}), 
using commutativity equations,  for metric
(\ref{metric0})
in the basis (\ref{basisf}) we obtain
\beaa
&&
(*F)_3=-\frac{1}{2}{\e}^{\phi}{m_t}
[\nabla_1,\nabla_2]
-\frac{1}{2}
\frac{m_{tt}}{m_t}
\nabla_3,
\\
&&
(*F)_2=[\nabla_3,\nabla_2],
\\
&&
(*F)_1=[\nabla_1,\nabla_3] - 
\left(\frac{m_x}{m_t}\right)_t
\nabla_3.
\eeaa
A vector field in the r.h.s. of the first expression
cancels due to the second equation of system
(\ref{gen2DTL}), 
in the second it is absent,
in the third it cancels identically.
Comparing the matrix part  of compatibility
equations
(\ref{comm1}-\ref{comm3})
with Yang-Mills-Higgs equations
(\ref{Ward}), 
we come to the conclusion that it represents
equations 
(\ref{Ward}) with a 1-form
\bea
&&
\omega=
-2\left(\frac{m_x}{m_t}\right)_t\mathbf{e}^1
- 
\left(
2\phi_t+\frac{m_{tt}}{m_t}
\right)\mathbf{e}^3
\\
&&
=
\left(
\frac{m_{tt}}{m_t}\frac{m_x}{m_t}-
2\phi_t\frac{m_x}{m_t}-2\frac{m_{xt}}{m_t}
\right)
\d x
-
\left(
2\phi_t+\frac{m_{tt}}{m_t}
\right)\d t,
\eea
which, up to a conformal gauge corresponding
to metric
(\ref{metric0}),
$\omega \rightarrow \omega + 2\d \ln m_t$,
coincides with 1-form (\ref{Weyl2}).
\begin{prop}
The matrix part of commutativity equations 
for Lax pair 
(\ref{Laxgauge}) 
represents Yang-Mills-Higgs equations (\ref{Ward})
on the background of Einstein-Weyl
geometry
(\ref{Weyl1}-\ref{Weyl2})
\end{prop}
Using the statements of the work \cite{DFK14}
that Einstein-Weyl geometry 
(\ref{Weyl1}-\ref{Weyl2}) connected with system
(\ref{gen2DTL}) is a local form of a generic
Lorentzian (or, in the complex analytic case,
generic complex analytic)
three-dimensional Einstein-Weyl geometry,
we come to the following conclusion:
\begin{prop}
There exist local coordinates such that 
Yang-Mills-Higgs equations (\ref{Ward}) 
on the background of real Lorentzian
(general complex analytic) Einstein-Weyl geometry
reduce to equations of commutativity 
of operators (\ref{Laxgauge}), so that 
the  scalar part of commutativity conditions
corresponds to system
(\ref{gen2DTL})
defining an Einstein-Weyl geometry 
(\ref{Weyl1}), (\ref{Weyl2}),
and the matrix (gauge) part gives 
the Yang-Mills-Higgs equations 
on the background of this Einstein-Weyl geometry.
\end{prop}
\section
{The case of an Einstein-Weyl geometry with
the Euclidean signature}
\subsection{A symmetric Lax pair and equations}
The technique needed to make a transition from
system
(\ref{gen2DTL}) to 
the system symmetric in variables
$x$, $y$,
describing an Einstein-Weyl geometry with
the Euclidean signature,
was developed in the work
\cite{LVB10}.
This transition is described by a rather
simple transformation
\bea
t=\tau + \mathrm{i}\mu,
\nn
\\
m=\tau - \mathrm{i}\mu,
\label{trans}
\eea
where $\tau$ is a new independent variable,
and
$\mu$ is a new dependent variable. 
This transformation is considered 
for Einstein-Weyl geometry 
(\ref{gen2DTL}), (\ref{Weyl1}), (\ref{Weyl2}) 
in the complex analytic case, and a real Euclidean
reduction is defined by the condition that 
$\phi$, $\mu$, $\tau$ are real and $x$, $y$ are
complex conjugated, $x=z$, $y=\Bar z$. 
To perform a transition it is convenient
to rewrite system
(\ref{gen2DTL}) 
in terms of differential forms,
\beaa
&&
\d (\e^{-\phi})_t\wedge\d x\wedge \d y
=\d m\wedge d\phi_y\wedge\d y,
\\
&&
\e^{-\phi}\d m_t\wedge\d x\wedge \d y
=-\d m\wedge \d m_y\wedge\d y,
\eeaa
Having in mind the matrix extension and
the Yang-Mills-Higgs equations, we will perform
the transformation on the level
of a Lax pair.

The transformation to a symmetric system arises 
naturally if one considers the symmetries of
linear operators and wave functions of the hierarchy,
analogous to the symmetry of Lax pair
(\ref{LaxToda})  
for a simpler case of equation
(\ref{d2DTL}).
We will give a brief description in terms of the hierarchy
below, and now we will move on to a symmetric analogue of 
Lax pair
(\ref{Laxgen2DTL}) 
and construct its matrix extension,
similar to matrix extension
(\ref{Laxgauge})
for Lax pair
(\ref{Laxgen2DTL}).

A symmetric Lax pair
\bea
&&
\partial_{z}\mathbf\Psi=L_1\mathbf\Psi,\quad L_1
=
(\lambda \e^\varphi u + v)\partial_\tau + 
((\varphi_\tau v - \varphi_z)- \lambda u \e^\varphi \varphi_\tau)
\l\partial_\lambda,
\nn\\
&&
\partial_{\bar z}\mathbf\Psi=L_2\mathbf\Psi,
\quad L_2
=
(-\frac{1}{\lambda} \e^\varphi \bar u + \bar v)\partial_\tau - 
((\varphi_\tau \bar v - \varphi_{\bar z}) + \frac{1}{\l}\bar u \e^\varphi \varphi_\tau)
\l\partial_\lambda,
\label{Laxsym}
\eea
where $\varphi=-\frac{1}{2}\phi$,
is obtained from Lax pair (\ref{Laxgen2DTL})
by a transformation (\ref{trans}) 
plus an extra twisting of spectral variable
$\lambda\rightarrow\e^{\varphi}\lambda$,
it possesses a symmetry
$L_2(\lambda)=\bar L_1(-\bar\lambda^{-1})$.
Here we have introduced the notations
\beaa
u=\frac{1}{1 + \mathrm{i}\mu_\tau},
\quad
\Bar u=\frac{1}{1 - \mathrm{i}\mu_\tau},
\quad
\bar v=\frac{\mathrm{i}\mu_{\Bar z}}{1+\mathrm{i}\mu_\tau},
\quad
v=
-\frac{\mathrm{i}\mu_{z}}{1-\mathrm{i}\mu_\tau}.
\eeaa
Equations of commutativity of the Lax pair read
\beaa
&&
(v_{\bar z} \Red{-} \e^\varphi u \partial_\tau (\e^\varphi \bar u) + v \partial_\tau \bar v)
-\text{c.c.}=0,
\\
&&
(\partial_{\bar z}(\varphi_\tau v - \varphi_z) 
\Red{-} 
\e^\varphi u \partial_\tau(\bar u \e^\varphi\varphi_\tau)
-
v\partial_\tau(\varphi_\tau\bar v - \varphi_{\bar z}) 
{-}
u \bar u \e^{2\varphi}\varphi_\tau\varphi_\tau)
\nn\\&&\qquad\qquad
+\text{c.c.}=0,
\eeaa
or, in explicit form,
\bea
&&
\frac
{\e^{2\varphi}}
{1+\mu_\tau^2}
\mu_{\tau \tau}
+
\mu_{z\Bar z}
+
\frac
{\mu_{z}\mu_{\Bar z}\mu_{\tau \tau}}
{1+\mu_\tau^2}
-
\frac{\mathrm{i}\mu_{\Bar z}\mu_{\tau z}}{1+\mathrm{i}\mu_\tau}
+
\frac
{\mathrm{i}\mu_{z}\mu_{\tau \Bar z}}
{1-\mathrm{i}\mu_\tau}=0,
\label{eq1}
\\
&&
\frac{1}{2}
\frac
{(\e^{2\varphi})_{\tau\tau}}
{1+\mu_\tau^2}
+
\varphi_{z\Bar z}
+
\frac
{\mu_{z}\mu_{\Bar z}\varphi_{\tau \tau}}
{1+\mu_\tau^2}
+
\frac{\mathrm{i} \mu_z\varphi_{\tau {\Bar z}}}
{1-\mathrm{i}\mu_\tau}
-
\frac{\mathrm{i} \mu_{\Bar z}\varphi_{\tau {z}}}
{1+\mathrm{i}\mu_\tau} + F \varphi_\tau=0,
\label{eq2}
\\&&\qquad
2F=\left(
\frac
{\mu_{z}\mu_{\Bar z}}
{1+\mu_\tau^2}
\right)_\tau
+
\left(
\frac{\mathrm{i} \mu_z}
{1-\mathrm{i}\mu_\tau}
\right)_{\Bar z}
-
\left(
\frac{\mathrm{i} \mu_{\Bar z}}
{1+\mathrm{i}\mu_\tau}
\right)_{z}.
\nn
\eea
These equations can be obtained directly from system
(\ref{gen2DTL}) 
by  transformation
(\ref{trans}),
also giving the corresponding
Einstein-Weyl structure,
which we will write down below considering
the matrix extension.
\subsubsection*{Reductions}
Let us consider characteristic reductions 
of system
(\ref{eq1}-\ref{eq2}).

A Hamiltonian reduction for vector fields of 
Lax pair (\ref{Laxsym}) corresponds to
$\mu=0$ ($u=1$, $v=0$), equation (\ref{eq1})
vanishes, equation (\ref{eq2}) 
reduces to dispersionless
(2+1)-dimensional Toda equation
(\ref{d2DTL})
for $\phi=(-2\varphi)$.

A linearly degenerate case for which
vector fields of 
the Lax pair do not contain a derivative 
over the spectral variable corresponds to
$\varphi=0$, equation (\ref{eq2})
vanishes, from equation (\ref{eq1})
we get
\bea
(1+ \mu_{z}\mu_{\Bar z})\mu_{\tau \tau}
+ ({1+\mu_\tau^2})\mu_{z\Bar z}
-
\mathrm{i}({1-\mathrm{i}\mu_\tau})
{\mu_{\Bar z}\mu_{\tau z}}
+
\mathrm{i}({1+\mathrm{i}\mu_\tau})
{\mu_{z}\mu_{\tau \Bar z}}=0
\label{degen}
\eea

An interpolating reduction, which for system
(\ref{gen2DTL}) looks like 
$\phi=\beta\ln m_t$ \cite{LVB10}, 
\cite{LVB12},
for system
(\ref{eq1}-\ref{eq2}) takes the form
\beaa
\varphi=\beta \tan^{-1} \mu_\tau
\eeaa
and leads to the equation
\beaa
(\e^{\beta \tan^{-1} \mu_\tau}+ \mu_{z}\mu_{\Bar z})\mu_{\tau \tau}
+ ({1+\mu_\tau^2})\mu_{z\Bar z}
-
\mathrm{i}({1-\mathrm{i}\mu_\tau})
{\mu_{\Bar z}\mu_{\tau z}}
+
\mathrm{i}({1+\mathrm{i}\mu_\tau})
{\mu_{z}\mu_{\tau \Bar z}}=0
\eeaa
The limit
$\beta\rightarrow 0$ 
corresponds to the linearly-degenerate case
(\ref{degen}),
and $\beta\rightarrow \infty$  
to the Hamiltonian reduction (\ref{d2DTL}).
\subsection{A matrix extension}
Let us consider the matrix extension of Lax pair
(\ref{Laxsym})
\bea
&&
L_1
=
\partial_{z}-(\lambda \e^\varphi u + v)\partial_\tau - 
((\varphi_\tau v - \varphi_z)- \lambda u \e^\varphi \varphi_\tau)
\l\partial_\lambda + A - \lambda B,
\nn\\
&&
L_2
=
\partial_{\bar z}-
(-\frac{1}{\lambda} \e^\varphi \bar u + \bar v)\partial_\tau + 
((\varphi_\tau \bar v - \varphi_{\bar z}) 
+ \frac{1}{\l}\bar u \e^\varphi \varphi_\tau)
\l\partial_\lambda
+\wt A + \lambda^{-1} \wt B,\qquad
\label{ext0}
\eea
where $A$, $B$, $\wt A$, $\wt B$ 
are some matrix-valued functions.
We introduce a basis of vector fields
\beaa
\mathbf{e}_1=\partial_z-{v}\partial_{\tau},
\quad
\mathbf{e}_2=\partial_{\Bar z}-{\bar v}\partial_{\tau},
\quad
\mathbf{e}_3=\partial_\tau
\eeaa
and a dual basis of 1-forms
\beaa
\mathbf{e}^1=\d z,
\quad
\mathbf{e}^2=\d {\Bar z},
\quad
\mathbf{e}^3=
(\d\tau + v\d z + {\Bar v}\d {\Bar z}).
\eeaa
Let us rewrite  the matrix extension
of the Lax pair 
in terms of gauge covariant
vector fields
$\nabla_i=\nabla_{\mathbf{e}_i}=\mathbf{e}_i + A_i$
in the form
\bea
L_1=\nabla_1 - 
\lambda u \e^\varphi(\nabla_{3} - \mathrm{i}\Phi)
-((\varphi_\tau v - \varphi_z)- \lambda u \e^\varphi \varphi_\tau)
\l\partial_\lambda,
\nn\\
L_2=\nabla_2 + 
\frac{1}{\lambda}{\Bar u} \e^\varphi(\nabla_{3} + {\mathrm{i}}\Phi)
+((\varphi_\tau \bar v - \varphi_{\bar z}) 
+ \frac{1}{\l}\bar u \e^\varphi \varphi_\tau)
\l\partial_\lambda
\label{ext1}
\eea
First consider the trivial background case, 
$\varphi=0$,
$\mu=0$,
\beaa
L_1=\nabla_z - 
\lambda(\nabla_{\tau} - {\text{i}}\Phi)  
,
\\
L_2=\nabla_{\bar z} + 
\frac{1}{\lambda}(\nabla_{\tau} + {\text{i}}\Phi)
\eeaa
The conditions of commutativity of these
operators read
\bea
&&
[\nabla_z,\nabla_{\bar z}]=2{\text{i}}
[\nabla_{\tau},\Phi],
\nn\\
&&
[\nabla_z,\nabla_{\tau}]=
-\text{i}[\nabla_z,\Phi],
\nn\\
&&
[\nabla_{\bar z},\nabla_{\tau}]=
\text{i}[\nabla_{\bar z},\Phi],
\label{Bz}
\eea
or, in terms of real coordinates
$x^1$, $x^2$, $\tau$, $z:=x^1+\mathrm{i}x^2$,
\bea
&&
[\nabla_{x^1},\nabla_{x^2}]=
4
[\nabla_{\tau},\Phi],
\nn\\
&&
[\nabla_{x^1},\nabla_{\tau}]=
-[\nabla_{x^2},\Phi],
\nn\\
&&
[\nabla_{x^2},\nabla_{\tau}]=
[\nabla_{x^1},\Phi]. 
\label{B}
\eea
It is easy to see that equations 
(\ref{B}) represent the Bogomolny 
equations
\beaa
\mathrm{D}\Phi=*{F}
\eeaa
for the Euclidean metric
$g=4((\text{d}x^1)^2 + (\text{d}x^2)^2) + (\text{d}\tau)^2$,
and equations (\ref{Bz}) give a complex form
of these equations with the expression for the metric
$g=4\text{d}z\text{d}\bar z+ (\text{d}\tau)^2$.
\begin{remark}
Considering an imaginary coordinate
$\tau=\mathrm{i}t$ leads to Minkowski metric,
and equations
(\ref{Bz}) 
in a special gauge give a symmetric version
of the chiral model studied in
\cite{ZM81}.
\end{remark}

Let us proceed to the general case.
From the conditions of commutativity
$[L_1,L_2]=0$ 
for operators (\ref{ext1}) we get
\bea
&&
\frac{1}{2 \text{i}u{\Bar u}\e^{2\varphi}} 
[\nabla_1,\nabla_2] 
=
[\nabla_3,\Phi] 
+
2\varphi_\tau \Phi
+
\frac{1}{2 \mathrm{i}}
({\frac{\Bar u_\tau}{\Bar u}-\frac{u_\tau}{u}})\nabla_{3}
+
\frac{1}{2}
({\frac{\Bar u_\tau}{\Bar u}+\frac{u_\tau}{u}})\Phi,
\nn\\
&&
[\nabla_1, \nabla_3+\mathrm{i}\Phi]
+ \frac{{\Bar u}_z- v {\Bar u}_\tau}{\Bar u}
(\nabla_3+\mathrm{i}\Phi)=0
\nn\\
&&
[\nabla_2, \nabla_3 - \mathrm{i}\Phi]
+ \frac{{u}_{\Bar z}- {\Bar v} {u}_\tau}{u}
(\nabla_3-\mathrm{i}\Phi)=0,
\label{symcomm}
\eea
a complete system of commutativity equation
is obtained after adding equation
(\ref{eq2}), 
which corresponds to vanishing of the term 
with the derivative over a spectral variable
$\p_\lambda$.

A symmetric bivector of conformal structure
can be obtained from formula
(\ref{h0}) or from the symbol of linearisation
of system
(\ref{eq1}-\ref{eq2}), 
\bea
h=
(1+\mu_\tau^2)\e^{-2\varphi}
\mathbf{e}_1\cdot \mathbf{e}_2 
+ (\mathbf{e}_3)^2,
\label{h1}
\eea
the metric reads, respectively,
\bea
g=
4(1+\mu_\tau^2)^{-1} 
\e^{2\varphi}\mathbf{e}^1\cdot\mathbf{e}^2 + 
\mathbf{e}^3\cdot\mathbf{e}^3,
\label{metric1}
\eea
its signature is Euclidean.
Similar to the case of operators
(\ref{ext0}) 
it is possible to demonstrate that the matrix part
of commutativity conditions
(\ref{symcomm})
represents  Yang-Mills-Higgs system
(\ref{Ward})
\bea
\mathrm{D}\Phi+\tfrac{1}{2}\omega \Phi=*{F}
\label{Bgm}
\eea
for metric (conformal structure)
(\ref{metric1})
and 1-form
$\omega$ 
\bea
\omega= 
-
\left(
\frac
{\mathrm{i}\mu_{\Bar z}}
{1+\mathrm{i}\mu_\tau}
\right)_\tau 
\mathbf{e}^1
+
\left(
\frac
{\mathrm{i}\mu_{z}}
{1-\mathrm{i}\mu_\tau}
\right)_\tau 
\mathbf{e}^2
+\left(
4\varphi_\tau 
-\partial_\tau \ln(1+\mu_\tau^2)
\right)
\mathbf{e}^3,
\label{omega1}
\eea
or the Bogomolny equations on the background
of an Einstein-Weyl geometry.

Results of the work 
\cite{DFK14} 
imply that real Einstein-Weyl geometry
(\ref{metric1}),
(\ref{omega1}) 
connected with system
(\ref{eq1}-\ref{eq2})
gives a local form of a generic 
Einstein-Weyl geometry with Euclidean signature,
and we come to the following conclusion:
\begin{prop}
The Bogomolny equations on the background of
an Einstein-Weyl geometry
(real, with the Euclidean signature)
(\ref{Bgm}) 
locally reduce to equations of commutativity
of operators
(\ref{ext1}), 
so that the scalar part of commutativity
conditions (connected with vector fields)
corresponds to the system (\ref{eq1}-\ref{eq2}), 
defining an Einstein-Weyl geometry
(\ref{metric1}), (\ref{omega1}),
and the matrix (gauge) part gives the Bogomolny
equations on the background of this
Einstein-Weyl geometry.
\end{prop}

\section{Hierarchies, a matrix extension, the dressing
method}
In this part we will give a brief description 
of the hierarchies connected with systems
(\ref{gen2DTL}) and 
(\ref{eq1}-\ref{eq2}) 
and their matrix extensions.
Our presentation is based on the works
\cite{LVB10}, \cite{LVB12}.
Considering the hierarchies clarifies 
the connections between the systems and 
the origin of transformation 
(\ref{trans}).
\subsection{The generalised dispersionless d2DTL hierarchy
and its matrix extension} 

The set of Lax-Sato equations of the hierarchy
connected with system 
(\ref{gen2DTL}) 
reads \cite{LVB10}, \cite{LVB12} 
\bea
&&
\left(
\partial_{x_n}
-\left(\frac{\e^{n\Lambda}\l\p_\l\Lambda}
{\{\Lambda,M\}}
\right)^\text{out}_+\p_t
+ 
\left(\frac{\e^{n\Lambda}\p_t\Lambda}
{\{\Lambda,M\}}
\right)^\text{out}_+
\l\p_\l
\right)
\begin{pmatrix}
\Lambda\\
M
\end{pmatrix}=0,\qquad
\label{Hi1}
\\
&&
\left(
\partial_{y_n}
+\left(\frac{\e^{-n\Lambda}\l\p_\l \Lambda}
{\{\Lambda,M\}}
\right)^\text{in}_-\p_t
- 
\left(\frac{\e^{-n\Lambda}\p_t\Lambda}
{\{\Lambda,M\}}
\right)^\text{in}_-
\l\p_\l
\right)
\begin{pmatrix}
\Lambda\\
M
\end{pmatrix}=0,\qquad
\label{Hi2}
\eea
where a Poisson bracket is defined as
$\{f,g\}=\l(f_\l g_t-f_t g_\l)$, and we consider
formal series
\bea
&&
\Lambda^\text{out}=\ln\lambda+\sum_{k=1}^{\infty}l^+_k\lambda^{-k},\quad
\Lambda^\text{in}=\ln\lambda+\phi+\sum_{k=1}^{\infty}l^-_k\lambda^{k},
\nn
\\
&&
M^\text{out}=M_0^\text{out} + \sum_{k=1}^{\infty}m^+_k\e^{-k\Lambda^\text{out}},\quad
M^\text{in}=M_0^\text{in} + m_0 +\sum_{k=1}^{\infty}m^-_k\e^{k\Lambda^\text{in}},
\nn
\\
&&
M_0=t+\sum_{k=1}^{\infty}x_k\e^{k\Lambda}-
\sum_{k=1}^{\infty}y_k\e^{-k\Lambda},
\label{LMseries}
\eea
where $\lambda$ is a spectral variable,
$(\dots)_-$, $(\dots)_+$ 
are standard projectors of Laurent 
series to negative and non-positive 
powers respectively.
Usually we suggest that `out' and `in' 
components define the functions  
inside and outside the unit circle in the complex plane, 
with $\Lambda^\text{in}-\ln\lambda$, 
$M^\text{in}-M_0^\text{in}$ 
analytic in the unit disc and 
$\Lambda^\text{out}-\ln\lambda$, 
$M^\text{out}-M_0^\text{out}$ 
analytic
outside the unit disc and decreasing at infinity.
For a function on the complex plane, having a discontinuity on the unit circle, 
by `in' and `out' components we mean the 
function inside and outside the unit disc.
Lax-Sato equations are equivalent to
the generating relation
\be
(\{\Lambda,M\}^{-1}\d\Lambda\wedge \d M)^\text{out}=
(\{\Lambda,M\}^{-1}\d\Lambda\wedge \d M)^\text{in},
\label{genToda}
\ee
suggesting analyticity of the 2-form
\be
\omega=\{\Lambda,M\}^{-1}\d\Lambda\wedge \d M
\ee
in the complex plane.
The dressing scheme for hierarchy
(\ref{Hi2})
can be formulated in terms of two-component
nonlinear Riemann-Hilbert problem
on the unit circle
\bea
\begin{aligned}
&\Lambda^\text{out}=F_1(\Lambda^\text{in},M^\text{in}),
\\
&M^\text{out}=F_2(\Lambda^\text{in},M^\text{in}),
\end{aligned}
\label{Riemann}
\eea
Lax-Sato equations for the times
$x=x_1$, $y=y_1$ 
correspond to Lax pair
(\ref{Laxgen2DTL}).

To obtain a matrix extension of the hierarchy,
we introduce in addition a matrix Riemann-Hilbert
problem
\bea
\Psi^\text{out}=
\Psi^\text{in}R(\Lambda^\text{in},M^\text{in}),
\label{RHT}
\eea
the matrix-valued function $\Psi$ 
is normalised by 1 at infinity and analytic,
having no zeroes, inside and outside the unit circle,
\be
\Psi^\text{out}=
1+\sum_{n=1}^\infty \Psi^+_n(\mathbf{t})\lambda^{-n},
\quad
\Psi^\text{in}=
\sum_{n=0}^\infty \Psi^-_n(\mathbf{t})\lambda^{n}.
\label{Psiser1}
\ee
Unity normalisation of the function 
$\Psi$ accounts for fixing the gauge
corresponding to the operators
of the form
(\ref{Laxmatrix}).
General operators of matrix (gauge) extension
of the type 
(\ref{Laxgauge}) require 
unnormalised regular (bounded at infinity)
solutions of problem
(\ref{RHT}). 
Matrix Riemann problem implies analyticity
of the matrix-valued 3-form
\beaa
\Omega=\omega\wedge\d\Psi\cdot\Psi^{-1},
\eeaa
corresponding to an additional 
generating relation for the matrix extension
of the hierarchy,
\bea
(\omega\wedge\d\Psi\cdot\Psi^{-1})^\text{in}
=(\omega\wedge\d\Psi\cdot\Psi^{-1})^\text{out},
\label{genTodaM}
\eea
which gives Lax-Sato equations 
for series 
(\ref{Psiser1}),
defining their evolution on the background
produced by Lax-Sato equations
(\ref{Hi2}),
\begin{align}
&
{\partial_{x_n}}
\begin{pmatrix}
\Lambda
\\
M
\end{pmatrix}
=V^+_n(\lambda)
\begin{pmatrix}
\Lambda
\\
M
\end{pmatrix},\quad
{\partial_{y_n}}
\begin{pmatrix}
\Lambda
\\
M
\end{pmatrix}
=V^-_n(\lambda)
\begin{pmatrix}
\Lambda
\\
M
\end{pmatrix}
,
\nn\\
&
{\partial_{x_n}}
{\Psi}=\left(V^+_n(\lambda)-
((V^+_n(\lambda)\Psi)\cdot\Psi^{-1})^\text{out}_+\right)
\Psi,
\nn\\
&
{\partial_{y_n}}
{\Psi}=\left(V^-_n(\lambda)-
((V^-_n(\lambda)\Psi)\cdot\Psi^{-1})^\text{in}_-\right)
\Psi,
\label{LSmatrix}
\end{align}
where vector fields
$V^+_n(\lambda)$, 
$V^-_n(\lambda)$ 
are defined by
(\ref{Hi2}) 
and have the coefficients polynomial
respectively in 
$\lambda$ and $\lambda^{-1}$.
\subsection{The hierarchy and its matrix extension
for symmetric system
(\ref{eq1}-\ref{eq2})}
The formal series in this case 
look symmetric with respect to the involution
transposing zero and infinity
\cite{LVB10},
\bea
&&
\Lambda^\text{out}=\ln\lambda+\varphi
+\sum_{k=1}^{\infty}l^+_k\lambda^{-k},\quad
\Lambda^\text{in}=\ln\lambda-\varphi
+\sum_{k=1}^{\infty}l^-_k\lambda^{k},
\nn
\\
&&
M^\text{out}=M_0^\text{out} + \mathrm{i}\mu
+ \sum_{k=1}^{\infty}m^+_k\e^{-k\Lambda^\text{out}},\quad
M^\text{in}=M_0^\text{in} - \mathrm{i}\mu  
+\sum_{k=1}^{\infty}m^-_k\e^{k\Lambda^\text{in}},
\nn
\\
&&
M_0=\tau +\sum_{k=1}^{\infty}z_k\e^{k\Lambda}-
\sum_{k=1}^{\infty}(-1)^{k-1}{\Bar z}_k\e^{-k\Lambda},
\label{LMseries1}
\eea
series of this form are obtained 
from series of the form 
(\ref{LMseries})
by transformation
(\ref{trans}) 
with an additional twisting
of a spectral variable
$\lambda\rightarrow\e^{\varphi}\lambda$,
$\varphi=-\frac{1}{2}\phi$.
A real Euclidean reduction of an Einstein-Weyl
geometry corresponds to the symmetry
\bea
&&
\Bar L\left(-{\Bar \lambda}^{-1}\right)
=-L^{-1}(\lambda), \quad L=\e^\Lambda,
\nn\\
&&
\Bar{M}\left(-{\Bar\lambda}^{-1}\right)
=M(\lambda)
\label{symser} 
\eea

Generating relation (\ref{genToda}) 
is unchanged,
and Lax-Sato equations are modified,
\bea
&&
\left(
\partial_{z_n}
-\left(
\frac
{\l\{\Lambda,M\}_{(\l,{z_n})}}
{\{\Lambda,M\}}
\right)^\text{out}_+\p_\tau
+ 
\left(\frac{\{\Lambda,M\}_{(\tau,{z_n})}}
{\{\Lambda,M\}}
\right)^\text{out}_+
\l\p_\l
\right)
\begin{pmatrix}
\Lambda\\
M
\end{pmatrix}=0,\qquad
\nn\\
&&
\left(
\partial_{{\Bar z}_n}
-
\left(
\frac
{\l\{\Lambda,M\}_{(\l,{\Bar z_n})}}
{\{\Lambda,M\}}
\right)^\text{in}_-\p_\tau
+
\left(\frac{\{\Lambda,M\}_{(\tau,{\Bar z_n})}}
{\{\Lambda,M\}}
\right)^\text{in}_-
\l\p_\l
\right)
\begin{pmatrix}
\Lambda\\
M
\end{pmatrix}=0,\qquad
\label{Hi21}
\eea
Where the definition of projector $(\dots)_-$
is changed to give non-positive powers 
of the series,
$\{f,g\}_{(x,y)}:=f_x g_y-f_y g_x$ 
(two-dimensional Jacobian).
Lax-Sato equations (\ref{Hi21}) for $z=z_1$,
$\Bar z=\Bar z_1$ correspond to Lax pair
(\ref{Laxsym}).

Reduction (\ref{symser}) can be provided
using a symmetric Riemann-Hilbert problem
\bea
\Bar{R}(-\e^{\Bar \Lambda^\text{out}},\Bar M^\text{out})=
R(\e^{-\Lambda^\text{in}},M^\text{in}),
\label{RHT1}
\eea 

A matrix extension is constructed similar to the
preceding case.
To preserve a symmetric form of the series
with respect to an involution of zero and infinity
(needed to perform reductions), we do not fix
the normalisation and consider series of the form
\beaa
\Psi^\text{out}=
\sum_{n=0}^\infty \Psi^+_n(\mathbf{t})\lambda^{-n},
\quad
\Psi^\text{in}=
\sum_{n=0}^\infty \Psi^-_n(\mathbf{t})\lambda^{n}.
\eeaa
Extended Lax-Sato equations are analogous to
equations
(\ref{LSmatrix}), 
\begin{align}
&
{\partial_{z_n}}
\begin{pmatrix}
\Lambda
\\
M
\end{pmatrix}
=V^+_n(\lambda)
\begin{pmatrix}
\Lambda
\\
M
\end{pmatrix},\quad
{\partial_{{\Bar z}_n}}
\begin{pmatrix}
\Lambda
\\
M
\end{pmatrix}
=V^-_n(\lambda)
\begin{pmatrix}
\Lambda
\\
M
\end{pmatrix}
,
\nn\\
&
\{ {\partial_{z_n}}-V^+_n(\lambda) \}
{\Psi}
=
\bigl(( \{ {\partial_{z_n}}-V^+_n(\lambda) \} \Psi)
\cdot\Psi^{-1}\bigr)^\text{out}_+
\Psi,
\nn\\
&
\{{\partial_{{\Bar z}_n}}-V^-_n(\lambda)\}
{\Psi}
=
\bigl(( \{ {\partial_{{\Bar z}_n}}-V^-_n(\lambda) \} \Psi)
\cdot\Psi^{-1}\bigr)^\text{in}_-
\Psi,
\label{LSmatrix1}
\end{align}
taking into account a modification of vector
fields and projectors corresponding to equations
(\ref{Hi21}).
Lax-Sato equations (\ref{LSmatrix1}) for $z=z_1$,
$\Bar z=\Bar z_1$ correspond to Lax pair
(\ref{ext0}), (\ref{ext1}).

\subsection*{Acknowledgements}
The author is sincerely grateful to E.V. Ferapontov
for useful discussions.

\subsection*{Conflict of Interest} 
The author declares no conflicts of interest.


\end{document}